%% file: main_v2.tex
\documentclass[aps,prl,reprint,longbibliography,notitlepage,superscriptaddress,preprintnumbers]{revtex4-1}

\input{packages}

\input{defs}

\newcommand{\update}[1]{#1}

\begin{document}

\title{Gluon Saturation Effects in Exclusive Heavy Vector Meson Photoproduction}

\author{Jani Penttala}
\email{janipenttala@physics.ucla.edu}
\affiliation{
Department of Physics and Astronomy, University of California, Los Angeles, CA 90095, USA
}
\affiliation{Mani L. Bhaumik Institute for Theoretical Physics, University of California, Los Angeles, California 90095, USA}

\author{Christophe Royon}
\email{christophe.royon@cern.ch
}
\affiliation{Department of Physics and Astronomy, The University of Kansas, Lawrence KS 66045, USA
}

\begin{abstract}
We study exclusive $\jpsi$ and $\Upsilon$ photoproduction for proton and Pb targets in the high-energy limit, with the energy dependence computed using the linear Balitsky--Fadin--Kuraev--Lipatov and the nonlinear Balitsky--Kovchegov evolution equations. The difference between these two evolution equations can be directly attributed to gluon saturation physics. We find that for proton targets there is no difference between the two approaches at the energies of the currently available data, while for Pb targets in $\jpsi$ production the data shows a clear preference for the evolution with gluon saturation.
\end{abstract}

\maketitle

\textit{Introduction.}
One of the main discoveries of the HERA collider was the rise of the gluon density in electron--proton collisions at low values of Bjorken $x$, the momentum fraction of the interacting 
\update{parton} 
in the proton~\cite{H1:2015ubc}. 
This rapid increase in the gluon density would eventually violate the unitarity of the scattering,
and hence gluon recombination effects taming the growth are expected to become important,
leading to gluon saturation
at very low $x$~\cite{Gribov:1983ivg}. 
To see these new effects, low values of $x$ (reached using a highly energetic collider) or a dense object (such as a heavy ion) are needed. 

The interaction with this highly gluonic target can be described using the color-glass condensate (CGC) effective field theory~\cite{Gelis:2010nm,Weigert:2005us}.
In this framework, the energy dependence of the process is described perturbatively using the nonlinear Balitsky--Kovchegov (BK) equation~\cite{Balitsky:1995ub,Kovchegov:1999yj}
which incorporates saturation effects into the evolution of the gluonic target.
In the region where saturation effects can be neglected, the BK equation reduces to the linear Balitsky--Fadin--Kuraev--Lipatov (BFKL) equation~\cite{Lipatov:1976zz,Kuraev:1977fs,Balitsky:1978ic}.
Thus, differences in results using these two evolution equations can be taken as a sign of gluon saturation.

Saturation effects are expected to become important when the relevant momentum scales of the process are comparable to the saturation scale $\Qs$, an emergent scale 
describing the strength of these saturation effects in the target.
At the energies available at the current experiments, the saturation scale in protons is estimated to be small,
$\order{\SI{1}{GeV}}$,
making saturation effects difficult to measure.
In a nuclear environment, however, the situation is different as the saturation scale gets boosted 
\update{due to the higher hadronic density of the large
nucleus}
by a factor of $A^{1/3}$, where $A$ is the mass number of the nucleus. 
This nuclear enhancement of saturation means that it should be possible to probe saturation at the Large Hadron Collider (LHC), using the currently available nuclear data, and also at the future Electron--Ion Collider (EIC)~\cite{Accardi:2012qut,Aschenauer:2017jsk,AbdulKhalek:2021gbh}.

While hints of saturation have already been identified~\cite{Albacete:2014fwa,Morreale:2021pnn} at the Relativistic Heavy-Ion Collider (RHIC)~\cite{PHENIX:2011puq,STAR:2021fgw} and the LHC~\cite{ALICE:2012mj,LHCb:2021vww}, these effects can also be attributed to other sources.
To find definite evidence of the gluon saturation phenomenon,
it is thus important to study observables that are very sensitive to saturation.
One promising process is exclusive heavy vector meson photoproduction,
where the mass of the vector meson is comparable to the saturation scale but still large enough for the process to be perturbative.
Another advantage of this process is that it is roughly proportional to the gluon density squared, making it more sensitive to saturation than inclusive processes that depend on the gluon density only linearly~\cite{Ryskin:1992ui,Frankfurt:2022jns}.
Experimentally, the main advantage of photoproduction is that it can be measured at very high energies in ultra-peripheral collisions at the LHC with both proton and Pb targets~\cite{ALICE:2014eof,ALICE:2018oyo,LHCb:2014acg,LHCb:2018rcm,ALICE:2023jgu,CMS:2023snh}.

In this Letter, we present results for exclusive $\jpsi$ and $\Upsilon$ photoproduction with proton and Pb targets and demonstrate that especially $\jpsi$ production with nuclear targets is sensitive to gluon saturation.
While 
\update{
the gluon content of protons and nuclei has been studied extensively in the exclusive vector meson production also before,
both
in the dipole picture~\cite{Bautista:2016xnp,ArroyoGarcia:2019cfl,Hentschinski:2020yfm,Peredo:2023oym,Mantysaari:2018nng}
and in the collinear factorization framework~\cite{Kopeliovich:1991pu,Andronic:2015wma,Eskola:2022vaf,Eskola:2023oos,Flett:2020duk,Flett:2024htj,Flett:2019pux,Flett:2021fvo,Guzey:2013xba,Guzey:2020ntc},}
our setup allows for direct inference of saturation effects by comparing results using the linear BFKL and nonlinear BK equations.
Another crucial ingredient of our calculation is taking into account the impact-parameter dependence in the dipole amplitude, as neglecting it can lead to incorrectly overestimating the saturation effects in the BK equation (see Sec.~IV A in~\cite{Mantysaari:2024zxq}), resulting in huge differences between the BFKL and BK equations even when the saturation effects are expected to be small~\cite{Dumitru:2023sjd}.
Including the dependence on the impact parameter allows one to treat BFKL and BK equations on a more equal footing, making direct comparisons between the two equations possible.

\textit{Exclusive vector meson production in the dipole picture.}
In the high-energy limit of QCD, scattering processes can be computed using the dipole picture.
This allows us to factorize the process $\gamma + p/A \to V + p/A $ into three different parts.
First, the photon fluctuates into a quark--antiquark dipole, which can be described in terms of the photon light-cone wave function.
Second, the quark--antiquark dipole scatters off the target eikonally, given by the nonperturbative dipole amplitude~\cite{McLerran:1993ka,McLerran:1994vd,McLerran:1993ni}.
Third, the dipole forms the vector meson, described by the vector meson light-cone wave function.
The production amplitude can then be written in terms of these three different parts as~\cite{Kowalski:2006hc,Hatta:2017cte,Mantysaari:2020lhf,Marquet:2010cf}:
\begin{multline}
\label{eq:amplitude}
    -i \mathcal{A}_\lambda = \int \dd[2]{\bt} \dd[2]{\rt} \int_0^1 \frac{\dd{z}}{4\pi    }
    e^{-i \Deltat \vdot \qty( \bt + (z-\frac{1}{2} ) \rt )}
    \\ 
    \times
     N\qty(\bt + \frac{1}{2}\rt , \bt - \frac{1}{2} \rt, \xpom) 
    \psi_\lambda^{\gamma \to q \bar q}(\rt, z) \qty[\psi_\lambda^{V\to q \bar q}(\rt, z)]^*    
\end{multline}
where $ \psi^{\gamma \to q \bar q}$ and $ \psi^{V \to q \bar q}$ are the light-cone wave functions for the photon and meson, $N$ is the dipole amplitude, $\rt$ is the transverse size of the quark--antiquark dipole, $\bt$ is the impact parameter, $\Deltat$ is the transverse-momentum exchange with the target, $z$ is the longitudinal momentum fraction of the quark, and $\lambda$ is the polarization of the photon and the meson.
We neglect the case where the photon and meson have a different polarization as this is heavily suppressed~\cite{Mantysaari:2020lhf}.
The coordinates $\bt + \frac{1}{2}\rt$ and $\bt - \frac{1}{2}\rt$ in the dipole amplitude correspond to the transverse coordinates of the quark and the antiquark scattering off the target.
In the high-energy limit, the transverse-momentum exchange $\Deltat$ is related to the Mandelstam variable $t$ by $t = - \Deltat^2$, and the dipole amplitude depends on the variable $\xpom = (M_V^2 +Q^2 - t)/(W^2 + Q^2)$
where $M_V$ is the meson mass, $Q^2$ is the photon virtuality, and $W$ is the center-of-mass energy of the photon--target system.
In this work, we only consider photoproduction where $Q^2 \approx 0$, and
we can thus neglect the contribution from longitudinally polarized photons that is suppressed in this limit.
We will also focus only on coherent production where the target remains intact.
The differential cross section in this case can then be written as
\begin{equation}
    \frac{\dd{\sigma}^{\gamma + p/A \to V + p/A}}{\dd{t}} = \frac{1}{4\pi}  
    \frac{1}{2}\sum_{\lambda = \pm 1} \abs{ \mathcal{A}_\lambda}^2
    .
\end{equation}

The photon wave function $\psi^{\gamma \to q \bar q}$ can be calculated perturbatively whereas the meson wave function is a nonperturbative quantity that we will model using the Boosted Gaussian approach~\cite{Kowalski:2006hc}.
The overlap of the photon and meson wave functions for transverse polarization can then be written as~\cite{Kowalski:2006hc}
\begin{align}
    &\psi_T^{\gamma \to q \bar q}(\rt, z) \qty[\psi_T^{V\to q \bar q}(\rt, z)]^*
    = \frac{e_f e \nc}{\pi z(1-z)} 
    \\
 &\times   \qty{ m^2 K_0( m r )  - \qty[z^2 + (1-z)^2] m K_1(m r) \partial_r  } \phi(r, z) \nonumber
\end{align}
where $r = \abs{\rt}$, $m$ and $e_f$ are the mass and the fractional charge of the quark, and $\phi$ is the scalar part of the meson's wave function.
This overlap is the same for both transverse polarizations $\lambda = \pm 1$,
and the scalar part takes the form
\begin{multline}
    \phi(r,z) = \mathcal{N} z(1-z)\\
     \times \exp( - \frac{m^2 \mathcal{R}^2}{8z(1-z)} - \frac{2z(1-z)r^2}{\mathcal{R}^2} + \frac{m^2 \mathcal{R}^2}{2} )    
\end{multline}
where $\mathcal{N}$ and $\mathcal{R}$ are free parameters fixed by the normalization and the leptonic width of the meson.
For $\jpsi$ we follow~\cite{Mantysaari:2018nng} and use the values 
\begin{equation}
\begin{aligned}
M_{\jpsi}&= \SI{3.097}{GeV}
&
m_c&= \SI{1.3528}{GeV}\\
\mathcal{N} &= 0.5890
&
\mathcal{R} &= \SI{1.5070}{GeV^{-1}},    
\end{aligned}
\end{equation}
and for $\Upsilon$ we take 
\begin{equation}
\begin{aligned}
M_{\Upsilon}&= \SI{9.460}{GeV}
&
m_b&= \SI{4.6000}{GeV} \\
\mathcal{N} &= 0.4555
&
\mathcal{R} &= \SI{0.77227}{GeV^{-1}}.
\end{aligned}
\end{equation}
The bottom quark mass is chosen to reproduce the correct normalization of the $\Upsilon$ production cross section.

\begin{figure*}[t]
	\centering
    \begin{subfigure}[T]{0.32\textwidth}
        \centering
        \includegraphics[width=\textwidth]{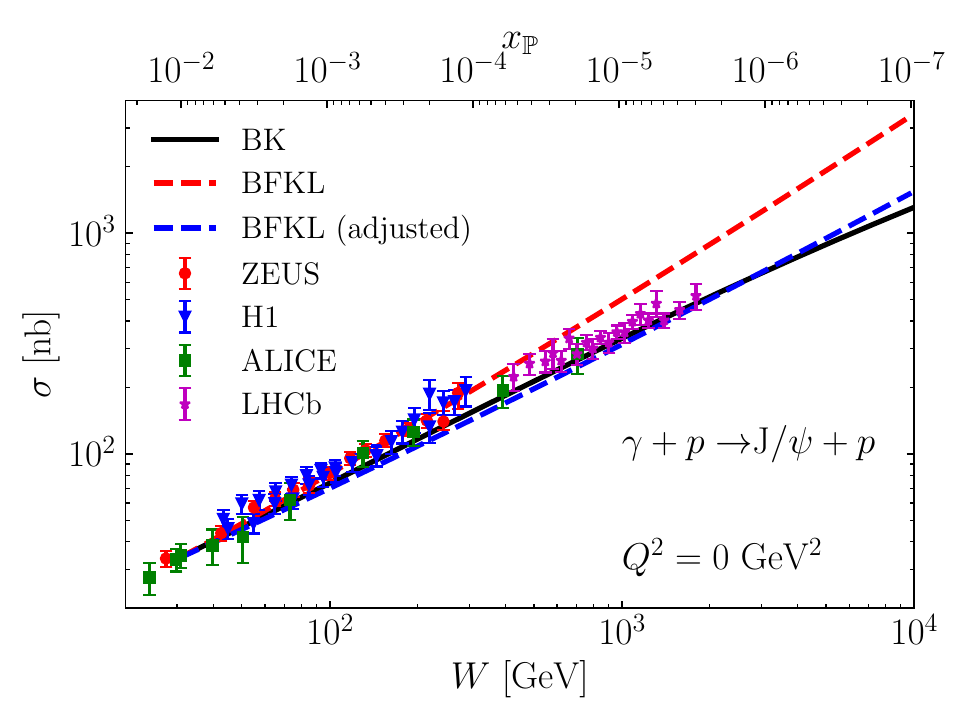}
        \caption{ Proton. }
        \label{fig:jpsi_W_p}
    \end{subfigure}
    \begin{subfigure}[T]{0.32\textwidth}
        \centering
        \includegraphics[width=\textwidth]{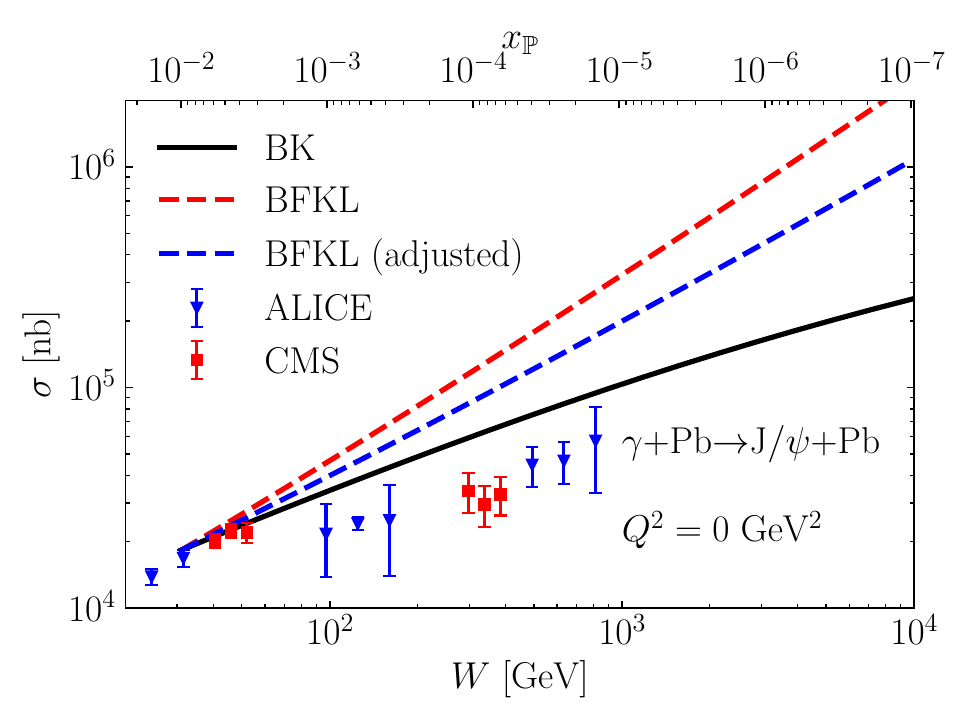}
        \caption{ Lead. }
        \label{fig:jpsi_W_Pb}
    \end{subfigure}
    \begin{subfigure}[T]{0.32\textwidth}
        \centering
        \includegraphics[width=\textwidth]{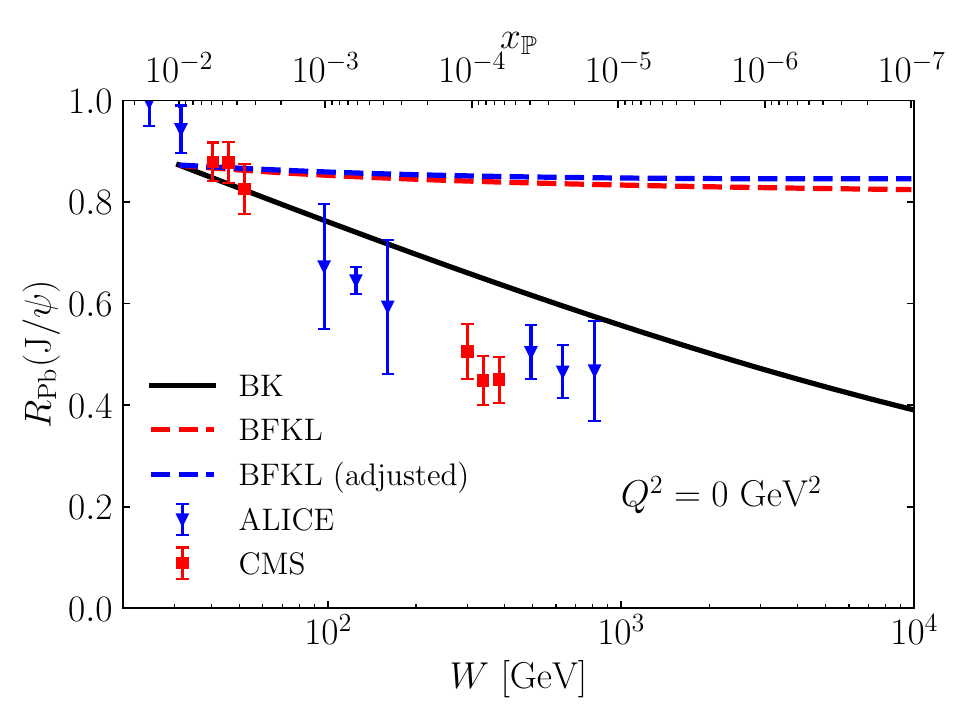}
        \caption{ Nuclear suppression factor. }
        \label{fig:jpsi_W_R}
    \end{subfigure}
    \caption{Exclusive $\jpsi$ photoproduction as a function of the center-of-mass energy $W$, using BFKL and BK evolution for the dipole amplitude. 
    For proton targets, we compare to the 
    experimental data from 
    ZEUS~\cite{ZEUS:2002wfj}, 
    H1~\cite{H1:2005dtp,H1:2013okq}, 
    ALICE~\cite{ALICE:2014eof,ALICE:2018oyo}, 
    and
    LHCb~\cite{LHCb:2014acg,LHCb:2018rcm}.
    For Pb targets and the nuclear suppression factor, the results are compared to the experimental data from
    ALICE~\cite{ALICE:2023jgu} and
    CMS~\cite{CMS:2023snh}.
    }
    \label{fig:jpsi_W}
\end{figure*}

\textit{High-energy evolution of the dipole amplitude.}
The dipole amplitude describing the interaction with the target 
is a nonperturbative quantity that has a perturbative evolution in terms of the energy of the scattering.
In this work, we consider two different evolution equations for the dipole amplitude: the BK and BFKL equations. 
These evolution equations resum large logarithms coming from subsequent emissions of slow gluons carrying only a small fraction of the emitting particle's longitudinal momentum.
The difference between the two equations is that the BFKL equation is completely linear in terms of the dipole amplitude, whereas the BK equation also includes a nonlinear term that takes into account gluon saturation.
We can write them compactly as
\begin{equation}
\begin{split}
        &\frac{\partial}{\partial \log 1/x} N(\xt_0, \xt_1, x)
        =  \int \dd[2]{\xt_2} \mathcal{K}(\xt_0, \xt_1, \xt_2) \\ 
        &\times 
        \bigl[  
        N\qty(\xt_0, \xt_2, x ) + N(\xt_1, \xt_2, x)
        - N(\xt_0, \xt_1, x) \\
        &- \sat N\qty(\xt_0, \xt_2, x )  N(\xt_1, \xt_2, x)
        \bigr]
\end{split}
\end{equation}
where the final term incorporating saturation effects is $\sat = 0$ for BFKL and $\sat = 1$ for BK.
The kernel $\mathcal{K}$ is given by
\begin{equation}
    \label{eq:evolution_kernel}
     \mathcal{K} (\xt_0,\xt_1,\xt_2) 
     = \frac{ N_c}{2\pi^2}
     \abs\Big{  \Kt(\xt_{20}) -
     \Kt(\xt_{21}) }^2 
\end{equation}
with the transverse vector
\begin{equation}
\label{eq:K_gluon_emission}
    \Kt^i(\xt) =\sqrt{\alpha_s \qty\big(\xt^2)}\times \frac{\mir 
 \xt^i}{\abs{\xt}}K_1(\mir \abs{\xt})
\end{equation}
where $\xt_{ij} = \xt_i - \xt_j$.
\update{
Here we have followed~\cite{Mantysaari:2024zxq} and adopted the \textit{daughter-dipole} prescription for the running of the coupling~\cite{Lappi:2012vw,Altinoluk:2023krt,Kovner:2023vsy}.
This choice makes it easier to compare the evolution and the parameters to computations done using the JIMWLK evolution, such as~\cite{Mantysaari:2022sux}, where many other running-coupling prescriptions, e.g.~\cite{Balitsky:2006wa}, are not possible.
We have also introduced the infrared regulator $\mir = \SI{0.4}{GeV}$ following~\cite{Schlichting:2014ipa,Mantysaari:2018zdd,Mantysaari:2024zxq}} to suppress nonperturbative effects in the evolution~\cite{Kovner:2001bh,Kovner:2002xa,Kovner:2002yt}.
This is required in the impact-parameter-dependent BK evolution to suppress so-called Coulomb tails, but it also makes the BFKL evolution more stable and allows for a more direct comparison of the two evolutions.
The coordinate-space running coupling constant in Eq.~\eqref{eq:evolution_kernel} is given as~\cite{Mantysaari:2024zxq}
\begin{equation}
    \alpha_s(r^2)
    = \frac{4 \pi}{\beta_0 
    \log \qty[   \qty(\frac{\mu_0^2}{\lambdaQCD^2})^{1 / \zeta}+
     \qty(\frac{4}{r^2\lambdaQCD^2})^{1 / \zeta}]^\zeta
    }
\end{equation}
where $\beta_0 = (11N_c - 2 N_f)/3 $, $N_f = 3$, and the parameters regulating the infrared region are chosen as
$\mu_0 = \SI{0.28}{GeV}$ and $\zeta= 0.2$~\cite{Lappi:2012vw,Mantysaari:2018zdd,Mantysaari:2022sux,Mantysaari:2024zxq}.
Our results are not sensitive to the infrared regulators $\mu_0$ and $\zeta$, but the parameter $\lambdaQCD$ controls the evolution speed.
As our standard setup, we follow~\cite{Mantysaari:2024zxq} and choose $\lambdaQCD = \SI{0.025}{GeV}$  that has been found to match the energy dependence of the data for BK evolution, and we note that this choice effectively takes into account the uncertainty from evaluating the strong coupling constant in the coordinate space instead of the momentum space~\cite{Mantysaari:2022sux}.

To compare the effects of saturation in the evolution, we use the same initial condition for both the BK and BFKL equations.
The initial condition for protons and nuclear targets is chosen as the impact-parameter-dependent McLerran--Venugopalan model from~\cite{Mantysaari:2024zxq} that already incorporates some saturation effects in the sense that the dipole amplitude in the initial condition cannot exceed unity.
The difference between protons and nuclei is in the thickness function describing the transverse density of the color charge, and we choose the same parametrizations as in~\cite{Mantysaari:2024zxq}.
The motivation for using this initial condition is that the results from BK and BFKL equations are expected to agree at large $x$, and using the same initial condition allows us to focus on the effects arising from the evolution where the saturation effects are taken into account.
Thus, any differences in the results for dipole amplitudes evolved with the BK and BFKL evolutions can be taken as coming from gluon saturation.

\textit{Numerical results.}
In Fig~\ref{fig:jpsi_W}, we show the predictions for exclusive $\jpsi$ photoproduction with proton and Pb targets along with the nuclear suppression factor, compared to the experimental data.
The nuclear suppression factor is defined as $ R_A = \sqrt{\sigma^A /\sigma^A_\text{IA}}$
where $\sigma^A$ is the $t$-integrated cross section for $\gamma + A \to V + A$ and
\begin{equation}
    \sigma^A_\text{IA} 
    = \frac{\dd{\sigma^{\gamma + p \to V + p}}}{\dd{t}} \Bigg|_{t=0} \times \int \dd{\abs{t}}\abs{F_A(t)}^2 
\end{equation}
refers to the impulse approximation~\cite{Chew:1952fca,Guzey:2013xba}.
The integrated form factor $\int \dd{\abs{t}} \abs{F_A(t)}^2$ has been calculated following the CMS collaboration~\cite{CMS:2023snh}
(see~\cite{Mantysaari:2023xcu} for the detailed explanation of the calculation).

\begin{figure*}[t!]
	\centering
    \begin{subfigure}[T]{0.32\textwidth}
        \centering
        \includegraphics[width=\textwidth]{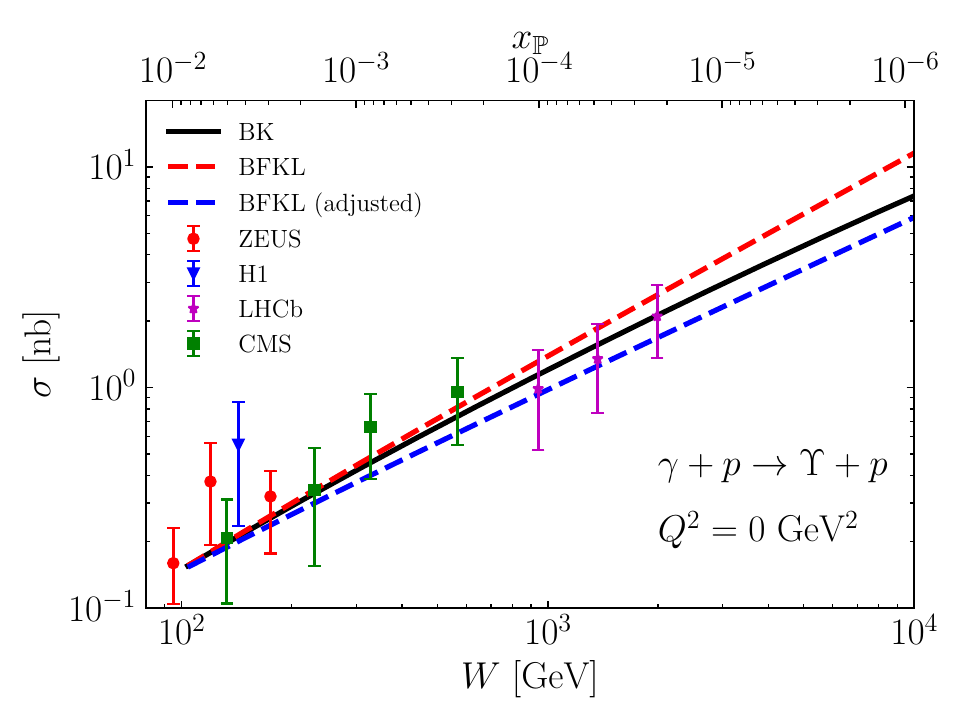}
        \caption{ Proton. }
        \label{fig:Upsilon_W_p}
    \end{subfigure}
    \begin{subfigure}[T]{0.32\textwidth}
        \centering
        \includegraphics[width=\textwidth]{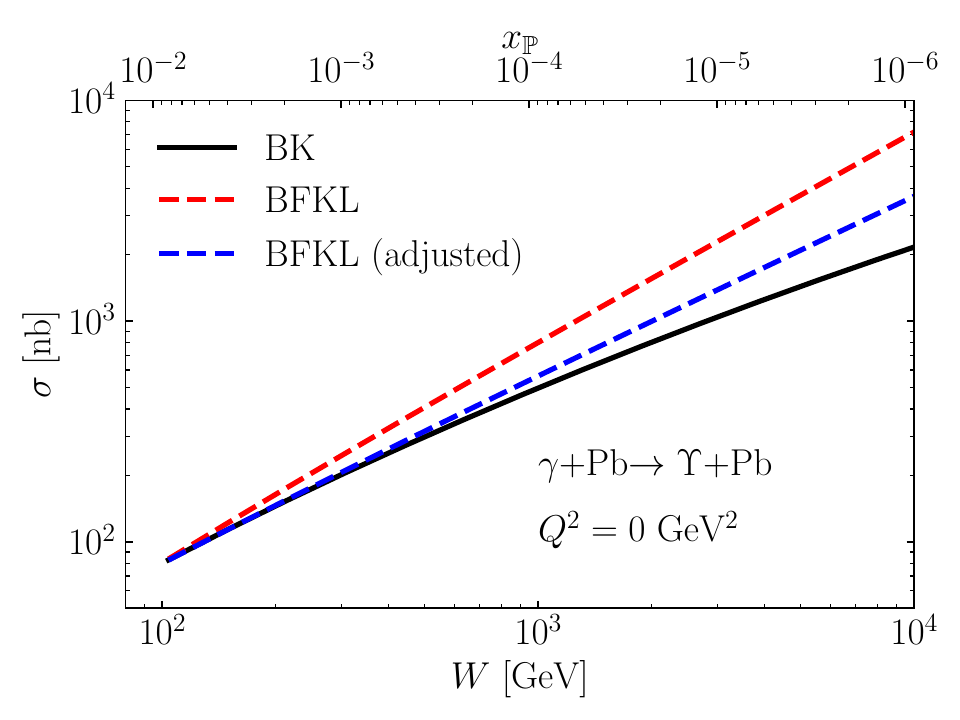}
        \caption{ Lead. }
        \label{fig:Upsilon_W_Pb}
    \end{subfigure}
    \begin{subfigure}[T]{0.32\textwidth}
        \centering
        \includegraphics[width=\textwidth]{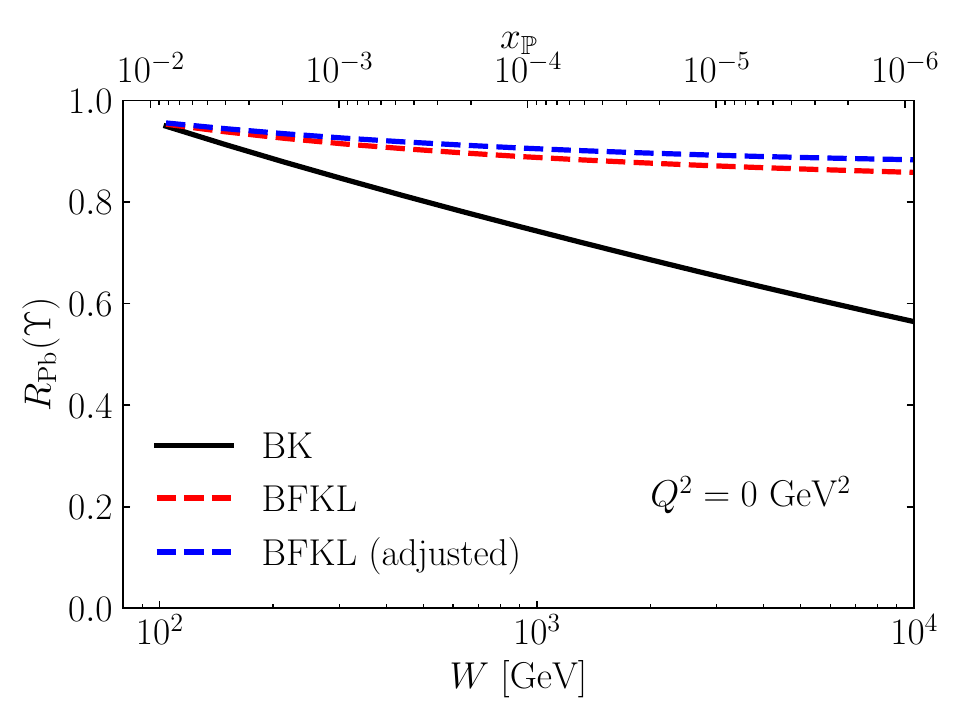}
        \caption{ Nuclear suppression factor. }
        \label{fig:Upsilon_W_R}
    \end{subfigure}
    \caption{Exclusive $\Upsilon$ photoproduction as a function of the center-of-mass energy $W$, using BFKL and BK evolution for the dipole amplitude. 
    For proton targets, we compare to the 
    experimental data from  ZEUS~\cite{ZEUS:1998cdr,ZEUS:2009asc},
    H1~\cite{H1:2000kis},
    CMS~\cite{CMS:2018bbk},
    and
    LHCb~\cite{LHCb:2015wlx}.
    }
    \label{fig:upsilon_W}
\end{figure*}

In addition to the BFKL results evolved with the same parameters as the BK setup, we also show adjusted BFKL results where the evolution speed has been tuned to the $\jpsi$ production data from proton targets by setting $\lambdaQCD = \SI{0.01}{GeV}$.
This is done to give a more fair comparison between the BK and BFKL results, as although the nonlinear effects in the evolution are not strong enough to alter the linear behavior of the energy dependence for proton targets, they slightly reduce the evolution speed.
This same value of $\lambdaQCD$ is then used for all adjusted BFKL results to be consistent with the results of $\jpsi$ production.
\update{
As studied in~\cite{Mantysaari:2024zxq}, 
when the evolution speed is fixed to the proton data, the results with nuclear targets 
are fairly robust under changes to the evolution such as altering the value of the infrared regulator $\mir$.}
Thus, the proton data gives strict constraints for our predictions for the nuclear data shown in Fig.~\ref{fig:jpsi_W_Pb}.
The data clearly disfavors the linear behavior of the BFKL predictions and is much closer to the BK results.
The same linear behavior of the BFKL is expected for both nuclei and protons, which follows from its linearity in terms of the dipole amplitude.
The similarity between the proton and nuclear results for BFKL is better seen in Fig.~\ref{fig:jpsi_W_R} where we show the nuclear suppression factor.
This measures the differences between proton and nuclear results and, due to the linear evolution, the BFKL predictions are roughly independent of the center-of-mass energy.

Predictions for exclusive $\Upsilon$ production are shown in Fig.~\ref{fig:upsilon_W}.
Again, the BK and BFKL results agree quite well for proton targets, and for nuclear targets they start to disagree only at much higher energies (or smaller values of $\xpom$) compared to $\jpsi$ production. 
This is because the relevant dipole sizes $ r \sim 1/M_V$ are larger for $\jpsi$ production, and saturation effects are generally more pronounced for larger dipoles.
Based on these results, we do not expect to see considerable effects of saturation in future measurements of exclusive $\Upsilon$ production at the LHC, and the measured data will be expected to agree with both BK and BFKL approaches.
This highlights the importance of measuring both exclusive $\jpsi$ and $\Upsilon$ production as the differences in the energy dependence would tell us about the onset of gluon saturation.

\textit{Discussion.}
We have computed the exclusive $\jpsi$ and $\Upsilon$ photoproduction in the high-energy limit where the interaction with the target nucleus can be described by the dipole amplitude.
The energy dependence of the dipole amplitude has been computed using both the linear BFKL and nonlinear BK evolution equations which allows us to estimate the importance of the gluon saturation.
Comparisons to the experimental data with nuclear targets indicate a strong preference for the results with the BK evolution that include saturation effects.
The linearity of the BFKL equation results in the energy dependence of the cross section being close to a simple power law,
$ \sigma \propto W^\delta $,
where the constant $\delta$ describes the evolution speed and should be roughly independent of the target.
This agrees with our finding that the nuclear suppression factor is roughly independent of the energy, in clear disagreement with the experimental data.

While our results are calculated only at the leading logarithmic accuracy for the evolution,
the energy dependence is expected to be similar also at higher orders.
Currently, neither the NLO BK nor BFKL have been implemented with the full impact-parameter dependence, which is the reason why we are working only at the leading logarithmic accuracy for the evolution.
We have checked that using the NLO BFKL equation including the collinear improvement~\cite{Ciafaloni:1999au,Ciafaloni:1999yw,Ciafaloni:2003rd}, but without the impact-parameter dependence, results in similar predictions. 
However, we emphasize that to compute exclusive vector meson production one needs the full impact-parameter-dependent dipole amplitude~\cite{Mantysaari:2024zxq}, and for this reason we choose to work at the leading logarithmic accuracy that is also closer to the setup with the BK evolution.
While the LO BFKL equation generally suffers from a strong dependence on the infrared region that results in too rapid evolution, including the infrared regulator as in our BK evolution is enough to avoid this problem.
It would be preferable to solve this problem with the collinear improvements to the evolution equations instead of an infrared cutoff,
but it is currently not known how to implement this for the BK evolution in exactly the same way as in the BFKL evolution:
the collinear improvement relies on the Mellin space formulation of the BFKL evolution which cannot be done for the BK evolution
(although there are schemes mimicking this behavior by resumming transverse logarithms in different ways~\cite{Beuf:2014uia,Iancu:2015joa,Iancu:2015vea,Ducloue:2019ezk,Ducloue:2019jmy}).
However, once the evolution speed is matched to the proton data, the nuclear results are almost independent of the exact form of the BK evolution used~\cite{Mantysaari:2024zxq}.
This makes our predictions for the energy dependence in the heavy nucleus case robust.

Even with the BK equation incorporating saturation effects, our predictions for Pb targets still overestimate the data.
We note that the overall normalization of the results depends highly on the vector meson wave function and the heavy quark masses used in the calculation, along with the phenomenological real-part and skewness corrections~\cite{Kowalski:2006hc} that we neglect here,
but the energy dependence is not very sensitive to these choices~\cite{Kowalski:2006hc}.
However, their effect on the normalization of the proton and nuclear results is very similar, and having correct normalization for both at the same time is very difficult~\cite{Mantysaari:2022sux,Mantysaari:2023xcu,Mantysaari:2024zxq}.
Possible solutions to this problem could be higher-order effects on the impact factor or a different initial condition for the dipole amplitude.
So far, this process has been calculated at next-to-leading order only in the nonrelativistic limit~\cite{Mantysaari:2021ryb,Mantysaari:2022kdm} which is not very reliable for photoproduction~\cite{Lappi:2020ufv}, and thus we are not able to estimate the effect of including higher-order corrections.
However, we note that the data prefer even more suppression for heavy nuclei than we predict, indicating that we are underestimating saturation effects with our model.
Taking this into account, e.g. by increasing the saturation scale in the initial condition or changing how we go from the protonic to nuclear dipole amplitude, would mainly change the predictions using the BK equations but not the BFKL ones, making gluon saturation even more pronounced in exclusive vector meson production.
Thus, it would not change our qualitative observation about the disagreement of the data with the BFKL results and the importance of including gluon saturation in the evolution.

Finally, we note that the collinear factorization framework, which is complementary to the dipole picture, has also been used to study exclusive heavy vector meson photoproduction with results that are closer to our BK setup than the BFKL one~\cite{ALICE:2023jgu,CMS:2023snh}.
While the collinear factorization approach does not explicitly include any saturation effects, the energy dependence of the parton distribution functions (PDFs) is not predicted by the theory but instead fitted to the data.
Depending on the model, some saturation effects can be implicitly included in the PDF fits.
In contrast, the energy dependence of the observables is predicted in the dipole picture, allowing one to turn off saturation in the evolution and making studying gluon saturation more straightforward.

While the theoretical description of this process depends on many nonperturbative components that need to be modeled, such as the meson wave function and the initial condition for the dipole amplitude, our predictions for the energy dependence of the nuclear cross section are robust once the energy dependence has been fitted to the proton data.
As we have demonstrated in this Letter, the existing data for exclusive $\jpsi$ photoproduction with Pb targets shows a clear preference for the existence of gluon saturation.
Future measurements of exclusive $\Upsilon$ production with Pb targets, along with the measurements of exclusive vector meson production at the EIC for different nuclei, are expected to give an even clearer, quantitative, picture about the saturation effects in the high-energy limit.

\textit{Acknowledgments.}
We would like to thank Edmond Iancu, Heikki Mäntysaari, and Farid Salazar for discussions.
J.P is supported by the National Science Foundation under grant No. PHY-1945471, and by the U.S. Department of Energy, Office of Science, Office of Nuclear Physics, within the framework of the Saturated Glue (SURGE) Topical Theory Collaboration.

\bibliographystyle{JHEP-2modlong.bst}
\bibliography{refs}

\end{document}

%% file: packages.tex
\usepackage[utf8]{inputenc}
\usepackage{mathrsfs}
\usepackage{amssymb}	
\usepackage{amsmath}
\usepackage{graphicx}
\usepackage{subdepth}
\usepackage{physics}
\usepackage{xcolor}
\usepackage{subcaption}
\usepackage{siunitx}

\definecolor{lcolor}{rgb}{0.5,0,0}
\definecolor{citcolor}{rgb}{0,0.3,0.0}

\usepackage[breaklinks,colorlinks,urlcolor=blue,citecolor=citcolor,linkcolor=lcolor]{hyperref}

%% file: defs.tex
\newcommand{\xt}{\mathbf{x}}
\newcommand{\Kt}{\mathbf{K}}
\newcommand{\bt}{\mathbf{b}}
\newcommand{\rt}{\mathbf{r}}

\newcommand{\Deltat}{\boldsymbol{\Delta}}
\newcommand{\nc}{N_\text{c}}

\newcommand{\Qs}{Q_\text{s}}
\newcommand{\mir}{m_\text{IR}}
\newcommand{\lambdaQCD}{\Lambda_\text{QCD}}

\newcommand{\xpom}{x_\mathbb{P}}

\newcommand{\sat}{\delta_\text{sat}}

\newcommand{\jpsi}{\text{J}/\psi}